\documentclass[5p,number,sort&compress]{elsarticle}
\usepackage{amsmath,amssymb}
\usepackage{graphicx}
\usepackage[colorlinks=true]{hyperref}

\newcommand{\zero}{ {(0)} }
\newcommand{\one}{ {(1)} }
\newcommand{\two}{ {(2)} } 
\newcommand{\rme}{{\mathrm{e}}}
\newcommand{\rmi}{{\mathrm{i}}}
\newcommand{\rmd}{{\mathrm{d}}}
\newcommand{\Rb}{R_\mathrm{b}}
\newcommand{\br}{\boldsymbol{r}}

\newcommand{\rout}{\bgroup \color{red} \ULdepth=-.5ex \ULset}
\newcommand{\bout}{\bgroup \color{blue} \ULdepth=-.5ex \ULset}

\begin{document}

\title{
  \vspace*{-2em} {\small \hfill RIKEN-iTHEMS-Report-23} \\
  \vspace{1.5em}
  Deconfinement transition in the revolving bag model}

\author[TUS,RIKEN]{Kazuya~Mameda}
\author[TUS]{Keiya~Takizawa}

\address[TUS]{Department of Physics, Tokyo University of Science, Tokyo 162-8601, Japan}
\address[RIKEN]{RIKEN iTHEMS, RIKEN, Wako 351-0198, Japan}


\begin{abstract}
Based on the bag model, we revisit the deconfinement phase transition under rotation.
On top of the usual rotational energy for noninteracting particles, we perturbatively analyze the revolution effect of the hadron bag, i.e., of the potential confining quarks.
The revolution effect can be phenomenologically translated into the rotational correction to the QCD vacuum energy or the gluon condensate.
We demonstrate that if the revolution effect is (is not) taken into account, the transition temperature increases (decreases) as the angular velocity increased.
The `revolving bag model' provides a feasible explanation of the recent lattice simulations showing, contrary to effective models, that rotation favors the confined phase.
\end{abstract} 

\maketitle

\section{Introduction}
Motivated by the measurements of strong vorticities in heavy-ion collisions~\cite{STAR:2017ckg}, the thermodynamic properties of quantum chromodynamics (QCD) under rotation have been one of the fascinating topics in high-energy physics.
Various effective models and perturbative approaches predict that rotation promotes chiral symmetry restoration~\cite{Chen:2015hfc,Jiang:2016wvv,Chernodub:2016kxh,Chen:2022mhf,Chen:2023cjt} and deconfinement~\cite{Fujimoto:2021xix,Chernodub:2020qah,Braga:2022yfe,Chen:2020ath,Chen:2022smf,Azuma:2023pzr}.
On the other hand, lattice QCD simulations cannot be directly available for rotating systems, because of the sign problem~\cite{Yamamoto:2013zwa}.
Nevertheless, the simulation with an imaginary angular velocity could be extrapolated to that with a real angular velocity, as long as the finite system-size is properly taken into account~\cite{Chen:2022smf}.
Such analyses indicate, in conflict with the model prediction, that rotation favors confinement~\cite{Braguta:2020biu,Braguta:2021jgn,Braguta:2021ucr,Braguta:2022str}.
This result is also supported by a lattice simulation with the Taylor expansion in terms of real angular velocity~\cite{Yang:2023vsw}.

This discrepancy suggests missing pieces in effective models and perturbative approaches under rotation.
In usual arguments, the rotational energy of noninteracting particles plays the most essential role to determine the thermodynamics under rotation~\cite{Vilenkin:1979ui}.
In other words, it is assumed that there is no rotational modification in the interactions mediated by gluons, and model parameters mimicking them.
This could however be an oversimplification, as is for the magnetic response of QCD~\cite{Bali:2011qj,Bali:2012zg,Fukushima:2012kc,Fraga:2012fs,Kojo:2012js,Kojo:2013uua}.
Rather, rotation, unlike electromagnetic field, directly couples with arbitrary fields including gluon and even ghost~\cite{Kugo:1977zq,Kugo:1979gm,Gribov:1977wm,Zwanziger:1993dh}.
It is hence conceivable that under rotation, such overlooked ingredients revise the usual model prediction.

The purpose of this work is to phenomenologically verify the above scenario, based on the bag model~\cite{Chodos:1974je,Chodos:1974pn}, which incorporates the two fundamental features of QCD, i.e., confinement and asymptotic freedom.
In the bag model, the deconfinement transition is determined through the pressure competition involving the bag constant, i.e., the contribution from the QCD vacuum energy.
The value of the bag constant is characterized predominantly by the kinetic energy of quarks confined inside a hadron bag~\cite{Bhaduri:1988gc,greiner2007quantum}.
The most significant perspective is that when rotation is applied to the bag model, it affects not only the quarks but also the bag itself.
Then, the rotation (more precisely, revolution) of the confining potential modifies the energy spectra of the quarks, and consequently the bag constant.
This is a great advantage of the `revolving bag model', compared with other models where details of interactions are already integrated out and subsumed into model parameters (see Ref.~\cite{Jiang:2021izj} for an exception).
In this work, we show that the revolution effect largely changes the critical temperature.

\section{Pressures of noninteracting gases}
The first building block of our model is the pressures of noninteracting gases.
To evaluate them, we consider the rotating cylindrical system with the radius $R$ and rigid angular velocity $\boldsymbol{\omega} = \omega \hat{z}$.
Because of the relativistic causality constraint $\omega R\leq 1$, we need to keep $R$ finite.
Under the Dirichlet-type boundary conditions, the energy dispersions for massless quarks ($i=q$), gluons ($i=g$) and pions ($i=\pi$) are~\cite{Davies:1996ks,Ambrus:2015lfr,Ebihara:2016fwa} 
\begin{equation}
\label{eq:e}
 \epsilon_i= \sqrt{\bigl(\xi_{\nu_i,k}/R\bigr)^2 + p_z^2} -\omega \mu_i,
\end{equation}
where $p_z$ is the continuous momentum along the $z$-direction, and $\mu_i$ is the total angular momentum;
$\mu_q = \pm \frac{1}{2},\pm \frac{3}{2},\cdots $ and $\mu_g,\mu_\pi = 0,\pm 1,\pm 2,\cdots$.
We defined $\nu_q =|\mu_q|-\frac{1}{2}$, $\nu_g=\mu_g$ and $\nu_\pi=\mu_\pi$, and $\xi_{\nu,k}=\xi_{-\nu,k}$ is the $k$-th smallest zero of the Bessel function of the first kind, $J_\nu(\xi)$.
As long as $\omega R\leq 1$ is imposed, the above dispersion is lower-bounded by $\mu_q = \frac{1}{2}$, $\mu_g=0$ and $\mu_\pi=0$.
Besides, while $\mu_i$ and $-\mu_i$ are degenerate for $\omega = 0$, they are not for $\omega>0$.

\begin{figure}
  \begin{center}
    \includegraphics[width=0.95\columnwidth]{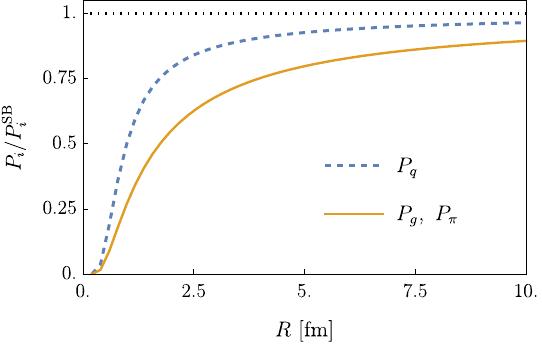}
    \caption{%
    Pressures of noninteracting gases for $\omega=0$ in the finite-size cylinder.
	}
  \label{fig:Ps}
  \end{center}
\end{figure}

In the standard manner, one straightforwardly derives the pressures.
The only difference is the transverse momentum phase space modified as~\cite{Ebihara:2016fwa}
\begin{equation}
\int \frac{\rmd p_x \rmd p_y}{(2\pi)^2} \to \frac{1}{\pi R^2} \sum_{\mu_i,k}.
\end{equation}
Thus the pressures become
\begin{equation}
\label{eq:Ps}
P_i 
= \eta_i\, d_i\, T \int \frac{\rmd p_z}{2\pi} \frac{1}{\pi R^2} \sum_{\mu_i,k} \ln \Bigl[1 + \eta_i\,\rme^{-\epsilon_i/T}\Bigr]
\end{equation}
with $\eta_q=1$, $\eta_g=\eta_\pi = -1$, $d_q = 2d_\mathrm{spin}N_\mathrm{f} N_\mathrm{c}$, $d_g=d_\mathrm{spin}(N_\mathrm{c}^2-1)$, $d_\pi = N_\mathrm{f}^2-1$ and $d_\mathrm{spin}=2$.
In this work, we take $N_\mathrm{f} = 2$ and $N_\mathrm{c}=3$, namely, $d_q=24$, $d_g=16$ and $d_\pi=3$.
The pressures~\eqref{eq:Ps} are also reproduced from the radial integration of the local expression in Ref.~\cite{Fujimoto:2021xix} (with discretized transverse momenta).
Figure~\ref{fig:Ps} shows the $R$-dependence of $P_i$'s at $\omega=0$ and $T=150\,\mathrm{MeV}$.
The pressures are normalized by the corresponding Stefan--Boltzmann (SB) limits: $(P_q^\mathrm{SB}, P_{g}^\mathrm{SB}, P_{\pi}^\mathrm{SB})=(7d_q/8,d_{g},d_{\pi})\times \pi^2T^4/90$.
At large $R$, we confirm the convergent behavior to the SB limit.
The deviations from 1 are caused by the finite-size effect. 

\section{Quarks confined in a revolving bag}
\begin{figure}
  \begin{center}
    \includegraphics[width=0.85\columnwidth]{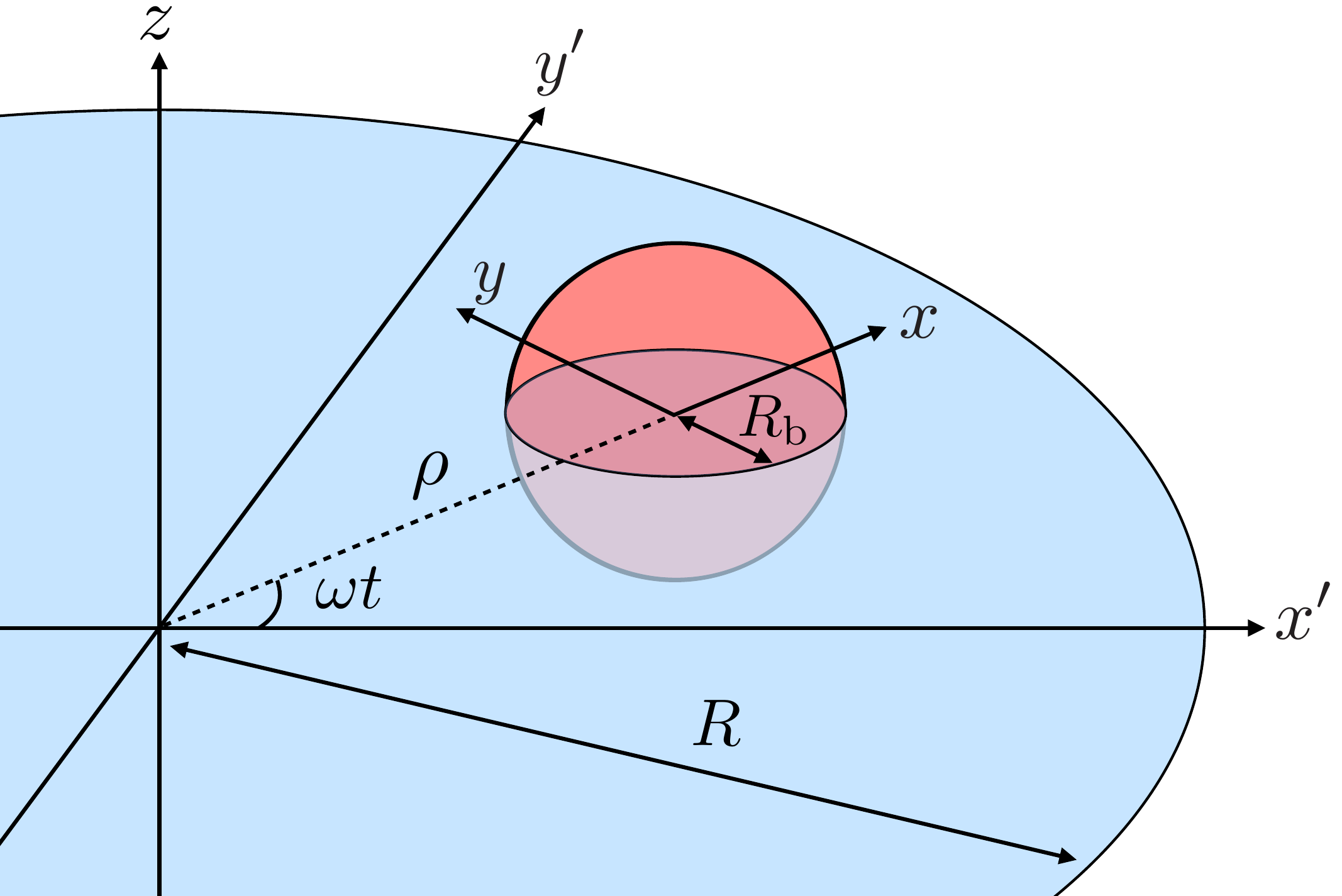}
    \caption{%
    Coordinate on the revolving spherical bag embedded in the cylinder.
	}
  \label{fig:bag}
  \end{center}
\end{figure}

Let us now study the rotational effect on the hadron bag.
For this purpose, we consider a spherical bag embedded into the cylindrical system, as depicted in Fig.~\ref{fig:bag}.
Their radii are $\Rb$ and $R$, respectively.
The bag revolves with angular velocity $\boldsymbol{\omega}=\omega\hat{z}$, so that it corotates with the above noninteracting gases.
We define $\rho\leq R$ as the difference between the bag center and the $z$-axis.
It is convenient to employ $x^\mu$ being the coordinate comoving with the bag, instead of the inertial coordinate $x'^\mu$.
We set $(x,y)$ and $(x',y')$ to be spanned on the same plane $z=0$, and $(x,y)=(0,0)$ is defined as the bag center.
Then, the coordinate transformation from $x'^\mu$ to $x^\mu$ is similar to that to the usual rigidly rotating coordinate.
Hence, one can adopt the following metric tensor $g_{\mu\nu}$ and vierbein ${e^a_\mu}$~\cite{Yamamoto:2013zwa}:
\begin{equation}
\begin{split}
 &g_{tt} = 1-\omega^2 (X^2+y^2),
 \quad g_{tx} = \omega y ,
 \quad g_{ty} = -\omega X , \\
 &e^0_t=e^1_x=e^2_y=e^3_z=1, \quad e^0_x=\omega y, \quad e^0_y=-\omega X 
\end{split}
\end{equation}
with $X:=x+\rho$.
Other components of $g_{\mu\nu}$ and ${e^a_\mu}$ are equivalent to the Minkowski metric $\eta_{\mu\nu}$ and zero, respectively.

We analyze the energy spectra of massless quarks confined inside the revolving bag.
In the bag coordinate $x^\mu$, the Dirac equation becomes $\rmi\partial_t \psi = (H_0+H_{1}+H_{2})\psi$ with
\begin{equation}
\label{eq:H01}
\begin{split}
 &H_0 = \gamma^0\bigl[-\rmi\, \boldsymbol{\gamma}\cdot\boldsymbol{\nabla} + U(r)\bigr], \quad
 H_{1} = - \omega J_z 
\end{split}
\end{equation}
and
\begin{equation}
\begin{split}
\label{eq:H2}
  H_{2} &=  \rmi\omega \rho\partial_y \\
 &\;= \sqrt{\frac{2\pi}{3}} \omega\rho
 	\biggl[
 		-(Y^{-1}_1+Y^{1}_1)\, \partial_r \\
 &\qquad
 		-\frac{1}{r}(Y^{-1}_1-Y^{1}_1) L_z
 		+\frac{1}{r}\frac{1}{\sqrt{2}}Y^{0}_1 
 		(L_+ + L_-)
 	\biggr].
\end{split}
\end{equation}
Here $U(r)$ is the infinite spherical potential well at $r=\Rb$ with $r=|\boldsymbol{r}|$ and $\boldsymbol{r}=(x,y,z)$.
This imitates the confining potential.
The angular momentum operators are $\boldsymbol{J} = \boldsymbol{L} + \boldsymbol{S}$, $\boldsymbol{L} = -\rmi\,\br\times \boldsymbol{\nabla}$, $\boldsymbol{S} = \frac{\rmi}{2} \boldsymbol{\gamma}\times\boldsymbol{\gamma}$, and $L_\pm = L_x\pm \rmi L_y$.
Both $H_{1}$ and $H_{2}$ are regarded as the rotational energy.
While the former is from the angular momentum $J_z$, the latter is due to $-\rmi\rho\partial_y$.
We emphasize that without $U(r)$, the latter could be eliminated by the translation along the $x$-axis.
In this model, hence, $H_{2}$ represents the bag revolution effect, or the rotational correction to the confining potential.

For $\omega = 0$, the Dirac equation is analytically solved~\cite{Chodos:1974pn}
(see also Refs.~\cite{Bhaduri:1988gc,greiner2007quantum}).
In the Dirac representation, the positive-energy solution is $\psi^\zero= \rme^{-\rmi E^\zero t} \varphi^\zero(\br)$ with $E^\zero=p$ and
\begin{equation}
\label{eq:varphi}
\begin{split}
 &\varphi^\zero(\br) = \frac{1}{\mathcal{N}}
 \begin{pmatrix}
 j_\ell(p r) \chi_{\kappa,\ell,\mu}(\hat{\br}) \\ 
 \rmi\, \mathrm{sgn}(\kappa) j_{\ell'}(p r) \chi_{-\kappa,\ell',\mu}(\hat{\br})
 \end{pmatrix},
\end{split}
\end{equation}
where $j_\ell(pr)$ is the spherical Bessel function.
The factor $\mathcal{N}$ is the normalization constant to fulfill $\int \rmd^3 \br\, \varphi^{\zero\dag} \varphi^\zero = 1$.
The quantum numbers are defined as $\boldsymbol{J}^2\psi = j(j+1)\psi$, $J_z\psi = \mu\psi$, and $K\psi = -\kappa\psi$ with $K = \gamma^0 (\boldsymbol{L}\cdot\boldsymbol{S} + 1)$, $j= \frac{1}{2},\frac{3}{2},\cdots$, $\mu = \pm \frac{1}{2},\pm\frac{3}{2},\cdots,\pm j$ and $\kappa = \pm (j+\frac{1}{2})$.
We introduced $\ell = \ell'+1 = j + \frac{1}{2} = \kappa$ for $\kappa>0$ and $\ell+1 = \ell' =  j + \frac{1}{2} = -\kappa$ for $\kappa<0$.
Also $\chi(\hat{\br})$ with $\hat{\br}=\br/r$ is the spinor spherical harmonics:
\begin{equation}
\begin{split}
 &\chi_{\kappa,\ell,\mu}(\hat{\br}) = 
 \frac{1}{\sqrt{2\ell+1}}
 \begin{pmatrix}
   \mathcal{A}_{\kappa,\ell,\mu}\, Y_\ell^{\mu-\frac{1}{2}}(\hat{\br}) \\
  \mathcal{B}_{\kappa,\ell,\mu}\, Y_\ell^{\mu+\frac{1}{2}}(\hat{\br})
 \end{pmatrix},
\end{split}
\end{equation} 
where the coefficients are $-\mathcal{A}_{\kappa>0,\ell,\mu}= \mathcal{B}_{\kappa<0,\ell,\mu} = \mathcal{C}^-_{\ell,\mu}$ and $\mathcal{A}_{\kappa<0,\ell,\mu} = \mathcal{B}_{\kappa>0,\ell,\mu} = \mathcal{C}^+_{\ell,\mu}$ with $\mathcal{C}_{\ell,\mu}^\pm=\sqrt{\ell\pm \mu+1/2}$.
Under the MIT boundary condition, the radial momentum is discretized as $p=\zeta_{k,\kappa,j}/\Rb$ with $\zeta=\zeta_{k,\kappa,j}$ being the $k$-th smallest solution of
\begin{equation}
\label{eq:MIT}
 j_\ell (\zeta) = -\mathrm{sgn}(\kappa) j_{\ell+1} (\zeta) .
\end{equation}
The energy dispersion reads
\begin{equation}
\label{eq:E0}
 E_N^{(0)}
 =  \frac{\zeta_n}{\Rb}, \quad N=\{n,\mu\}=\{k,\kappa,j,\mu\},
\end{equation}
which is degenerate in terms of $\mu$.
There are two ground states as
\begin{equation}
\label{eq:E0g}
	E_\mathrm{g}^{(0)}:= E_{N^{\pm}_{\mathrm{g}}}^{(0)} =\frac{\zeta_{n_\mathrm{g}}}{\Rb},\ \
	N_\mathrm{g}^{\pm}=\{n_\mathrm{g},\pm \tfrac{1}{2}\}
	=\{1,-1, \tfrac{1}{2}, \pm\tfrac{1}{2}\}.
\end{equation}

For $\omega \neq 0$, we need to implement a perturbation theory.
We can treat $H'=H_{1}+H_{2}$ as the time-independent perturbation, under the conditions $\omega \Rb \ll 1$ and $\omega\rho \leq \omega R\ll 1$.
The matrix element of $H'$ is
\begin{equation}
\label{eq:HMN}
 H'_{MN} = \int_{r\,\leq \Rb} \rmd^3 \br\, \varphi^{\zero\dag}_{M} H' \varphi^\zero_{N},
\end{equation}
where $\varphi^\zero_{N}$ is given by Eq.~\eqref{eq:varphi}.
This is evaluated with the help of the following angular integral formula~\cite{rose1995elementary}:
\begin{equation}
\begin{split}
\label{eq:three-Y}
&\int \rmd\Omega\, Y_{\ell_1}^{m_1} Y_{\ell_2}^{m_2}(Y_{\ell_3}^{m_3})^* \\
&=\biggl[\frac{(2\ell_1+1)(2\ell_2+1)}{4\pi(2\ell_3+1)}\biggr]^{1/2}\,
C^{\ell_1\ell_2\ell_3}_{000}
C^{\ell_1\ell_2\ell_3}_{m_1m_2m_3}
\end{split}
\end{equation}
with the Clebsch--Gordon coefficient
\begin{equation}
C^{\ell_1\ell_2\ell_3}_{m_1m_2m_3}=\bigl\langle {\ell_1}\,{m_1};{\ell_2}\,{m_2}\big|{\ell_1}\,{\ell_2};{\ell_3}\,{m_3}\bigr\rangle.
\end{equation}

Because of the degeneracy in Eq.~\eqref{eq:E0}, the energy corrections are calculated from the degenerate perturbation theory.
Let $N_1$ and $N_2$ be labels belonging to the same degenerate subspace, namely, $N_1=\{n,\mu_1\}$ and $N_2 =\{n,\mu_2\}$ for a given $n$.
The first-order energy correction is the eigenvalue of $H'_{N_1N_2}$.
From the selection rules imposed by Eq.~\eqref{eq:three-Y}, one finds that the matrix element is already diagonalized as $H'_{N_1N_2} = (H_{1})_{N_1N_2} = -\omega \mu_1 \delta_{\mu_1\mu_2}$.
Thus, we get
\begin{equation}
\label{eq:E1}
 E^\one_N = -\omega \mu .
\end{equation}
This is the usual rotational energy, and $E^\zero_N + E^\one_N$ becomes the exact energy dispersion when $\rho=0$.
One can prove that $E^\zero_N + E^\one_N > 0$ holds for arbitrary $N$, as long as $\omega \Rb\leq 1$ is imposed~\cite{Zhang:2020hct}.

The degeneracy is completely resolved up to the first-order perturbation.
Hence, the second-order energy correction reads
\begin{equation}
\label{eq:E2}
 E^\two_N 
 = \sum_{\bar n\neq n}\sum_{\bar{\mu}} \frac{H'_{N\bar{N}} H'_{\bar{N}N}}{E_N^\zero-E_{\bar{N}}^\zero}
\end{equation}
with $\bar{N}=\{\bar{n},\bar{\mu}\}$.
The above summation over all states for $\bar n\neq n$ excludes the diagonal element $(H_{1})_{\bar{N}N} = -\omega \mu\, \delta_{\bar{n}n}\delta_{\bar{\mu}\mu}$.
The energy correction for the two unperturbed ground states $N_\mathrm{g}^\pm$ is numerically computed as
\begin{equation}
\label{eq:E2g}
 E_{\mathrm{g}}^\two:=E_{N^\pm_\mathrm{g}}^\two = - \frac{\alpha}{\Rb}(\omega\rho)^2,
 \quad
 \alpha \simeq 1.07.
\end{equation}
The two states $N_\mathrm{g}^\pm$ receive the same second-order correction.
This is because $\omega^2$ does not distinguish the spin-up and spin-down,
while the first-order correction~\eqref{eq:E1} does.

\section{Bag constant}
In the bag model, the proton ground-state energy $E_p$ is described as
\begin{equation}
\label{eq:Ep}
 E_p =  E_q + E_\mathrm{other} + \frac{4\pi}{3}\Rb^3 b.
\end{equation}
Here $E_q $ is the kinetic energy of three ground-state quarks and $E_\mathrm{other}$ contains the interaction among quarks, the Casimir energy, the effect of meson clouds etc.
The last term is the bulk energy with the bag constant $b$.
For $\omega=0$, we identify $E_q= 3 E^\zero_{\mathrm{g}}$ and $E_p=M_p$ with the proton mass $M_p$, and parametrize $E_\mathrm{other}=Z/\Rb$ for the dimensional reason.

To generalize Eq.~\eqref{eq:Ep} to the finite $\omega$ case, we implement two assumptions.
First, since $E_q$ is more dominant than $E_\mathrm{other}$, it is likely that the same is true for their rotational corrections.
As the first attempt to pursue the revolution effect of the bag model, we here assume $E_\mathrm{other}$ to be unchanged under rotation.
Second, the spin-rotation coupling contributes to the ground-state energy of protons, unlike the orbit-rotation one%
\footnote{
Strictly speaking, because of $\omega R\leq 1$, the spectrum should become gapped as in Eq.~\eqref{eq:e}.
This is however irrelevant to the following analysis.
Indeed, the ground state is still the mode with zero orbital angular momentum, and the finite-size effect is of $O((M_p R)^{-2})$, which is negligible for our parameters leading to $M_p R=50$.
}.
It is then seems that we face the difficult question of how to construct the spin-$\frac{1}{2}$ proton with three spin-$\frac{1}{2}$ quarks and spin-$1$ gluons~\cite{Aidala:2012mv}.
However, a term linear in $\omega$ is prohibited from emerging in Eq.~\eqref{eq:Ep}, because $b$ is a pressure, a time-reversal even quantity.
Therefore, it is necessary that the spin-rotation coupling of the proton are totally canceled by that of quarks, i.e., Eq.~\eqref{eq:E1} (and by that from $E_\mathrm{other}$, if any).
Although the underlying physics of the cancelation is elusive in the present phenomenology, we just suppose it.

Eventually, the only relevant modification in Eq.~\eqref{eq:Ep} is the second-order energy correction~\eqref{eq:E2g}.
Replacing $E^\zero_\mathrm{g}$ with $E^\zero_\mathrm{g}+E^\two_\mathrm{g}$, we arrive at
\begin{equation}
\label{eq:Mp}
 M_{p} 
 = \frac{C-3\alpha(\omega\rho)^2}{\Rb}
 + \frac{4\pi}{3}\Rb^3 b .
\end{equation}
We introduced the parameter $C=3\zeta_{n_\mathrm{g}}+Z>0$, which we will choose so that an appropriate critical temperature is reproduced for $\omega=0$.
Minimizing $M_p$ with respect to $\Rb$, we identify $b=b(\rho)$, which is the bag pressure balanced by the internal energy.
Up to $O((\omega\rho)^2)$, this is
\begin{equation}
\label{eq:b}
 b (\rho)
 = B_0\biggl[1+\frac{9\alpha}{C}(\omega\rho)^2 \biggr] ,
\end{equation}
where $B_0 := C/(4\pi R_{\mathrm{b}0}^4)$ with $R_{\mathrm{b}0}:=4C/(3M_p)$ is the bag constant for $\omega=0$.

For later convenience, we also derive the global bag constant $B$ from $b(\rho)$.
Then, we need the radial distribution of the bag inside the cylinder with radius $R$.
In the simplest case with the homogeneously distributed bag, the bag constant is
\begin{equation}
\label{eq:B}
 B=\frac{1}{\pi R^2}\int_0^R 2\pi\rho\,\rmd \rho\, b(\rho)
 = B_0\biggl[1+\frac{9\alpha}{2C}(\omega R)^2 \biggr] .
\end{equation}
This is a presumable expression because $\omega R$ is the quantity to characterize the strength of rotation.
If the bag is distributed inhomogeneously, $B$ is of course modified.
Nevertheless, owing to the centrifugal force, such a distribution should be a monotonically increasing function of $\rho$.
As a result, the correction in $B$ should become larger.
In other words, Eq.~\eqref{eq:B} is the form involving the minimal revolution effect of the bag, and we utilize it in the following analysis.
Also we demonstrate that even such a minimal effect drastically affects the phase transition.

By the definition of $B$, Eq.~\eqref{eq:B} hints that the QCD vacuum is affected by rotation, in sharp contrast with the vacuum of noninteracting systems~\cite{Ebihara:2016fwa}.
It is difficult to validate this fact within the bag model.
Focusing on energy spectra, however, we can phenomenologically examine the rotational response of the QCD vacuum.
In general, the rotational effect becomes visible even at zero temperature and zero density, if the rotational energy overcomes the infrared energy gap~\cite{Ebihara:2016fwa}.
Unlike in noninteracting systems, such a situation is realized in magnetized systems~\cite{Chen:2015hfc,Hattori:2016njk}, where the Landau quantization pushes down the gap~\cite{Chen:2017xrj}.
A similar argument could be applicable to the QCD vacuum, which is effectively viewed as a color-magnetic medium;
this interpretation explains asymptotic freedom and qualitatively illustrates the vacuum energy~\cite{Savvidy:1977as,Nielsen:1978rm,Nielsen:1978tr,Bhaduri:1988gc,huang1992quarks}.
Therefore, the strong interaction enables the QCD vacuum to receive the visible rotational effect.
The above picture matches the intuition that while the perturbative vacuum has nothing to rotate, the QCD vacuum is a nonempty state.

\section{Critical temperature}
In the bag model, the critical temperature $T_\mathrm{c}(\omega)$ is the solution of
\begin{equation}
\begin{split}
\label{eq:P_balance}
& P_\pi (T,\omega) = P_q (T,\omega) +  P_g (T,\omega) - B(\omega),
\end{split}
\end{equation}
where $P_i$ and $B$ are given by Eqs.~\eqref{eq:Ps} and~\eqref{eq:B}, respectively.
For simplicity, we ignored other subdominant excitations in the hadronic phase.
Instead of Eq.~\eqref{eq:P_balance}, one can determine $T_\mathrm{c}$ from the local pressure competition, as in Ref.~\cite{Fujimoto:2021xix}.
However, we have numerically checked that the qualitative behavior of $T_\mathrm{c}$ is irrelevant to the radial position, except for the vicinities of the center and the boundary edge.
For this reason, we focus on the global pressure competition~\eqref{eq:P_balance}.

\begin{figure}
  \begin{center}
    \includegraphics[width=0.95\columnwidth]{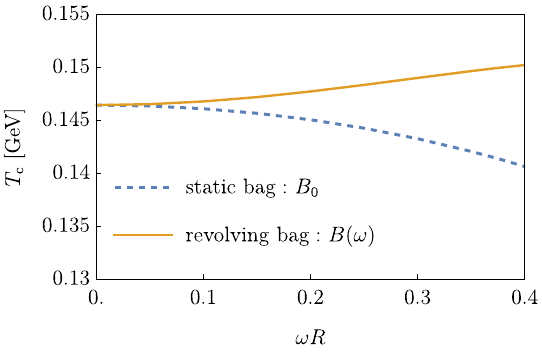}
    \caption{
    Critical temperature $T_\mathrm{c}$ as a function of angular velocity $\omega$ for the homogeneously distributed bag.
    }  \label{fig:Tc}
  \end{center}
\end{figure}

Figure~\ref{fig:Tc} shows the numerical solutions.
The solid line is $T_\mathrm{c}(\omega)$ from the revolving bag model with $B(\omega)$. 
The dash line is the result from the usual static bag, namely, the solution of Eq.~\eqref{eq:P_balance} under the replacement $B(\omega)\to B_0$.
We set $R=10\,\mathrm{fm}$, $M_p=1\,\mathrm{GeV}$ and $C= 2.51$, which lead to the standard value $B_0^{1/4}\simeq 200\,\mathrm{MeV}$.
For $\omega=0$, the solution is $T_\mathrm{c}(\omega = 0) \simeq 146\,\mathrm{MeV}$.
This is, due to the finite-size effect (see Fig.~\ref{fig:Ps}), slightly larger than that in the SB limit: $T_\mathrm{c}^\mathrm{SB} = [(\frac{7}{8}d_q+d_g-d_\pi)\pi^2/90]^{-1/4}B_0^{1/4}\simeq 144\,\mathrm{MeV}$.

For $\omega\neq 0$, the behavior of $T_\mathrm{c}$ strongly depends on whether the bag revolution is taken into account or not.
For the static bag with $B_0$, rotation decreases $T_\mathrm{c}$.
The underlying reason is clarified from the $\omega$-dependence of $P_i$.
Up to $O(\omega^2)$, the pressure correction $\Delta P_i := P_i(T,\omega)-P_i(T,0)$ is written as the familiar form of the rotational energy density of rigid body:
\begin{equation}
\label{eq:Delta_Ps}
 \Delta P_i= \frac{1}{2V}\omega^2 I_i ,
 \quad
  I_i := V\frac{\rmd^2 P_i(T,0)}{\rmd \omega^2}>0,
\end{equation}
where $I_i$ is the moment of inertia with the system volume $V=\pi R^2\int \rmd z$.
Also $I_\pi<I_q+I_g$ holds because of the large difference of $d_i$'s.
In Eq.~\eqref{eq:P_balance} with $\omega\neq 0$, hence, the net pressure correction $\Delta P_\mathrm{net}:=\Delta P_q+\Delta P_g-\Delta P_\pi>0$ should be compensated by the suppression of the temperature, i.e., $T_\mathrm{c} (\omega\neq 0)<T_\mathrm{c} (\omega=0)$.

The revolving bag with $B(\omega)$ yields the opposite behavior, namely, the enhancement of $T_\mathrm{c}$ by arbitrary $\omega\neq 0$.
In the revolving bag case, Eq.~\eqref{eq:P_balance} receives the modification not only from $\Delta P_\mathrm{net}$ but also from $\Delta B:=B(\omega)-B_0 = 9B_0\alpha(\omega R)^2/(2C)$.
From the same argument as before, one finds that $T_\mathrm{c} (\omega\neq 0)> T_\mathrm{c} (\omega=0)$ is a consequence of the competition $\Delta P_\mathrm{net}<\Delta B$.
The origin of $\Delta B$ is the rotational effect on the confining potential, which emulates an interaction mediated by gluons.
This is consistent with the following observation in Fig~\ref{fig:Tc};
the revolving (static) bag supports the lattice simulations (the effective approaches), where the rotational effect on the nonperturbative gluonic contribution is (is not) taken into account.

Another interpretation of our result is found from the relation between the bag constant and the QCD vacuum energy.
Let us rewrite $\Delta P_\mathrm{net}<\Delta B$ into
\begin{equation}
\label{eq:Ib}
 I_q+I_g - I_\pi < I_\mathrm{b}:=\frac{18\alpha}{C}\int \rmd V B_0 \rho^2 .
\end{equation}
The above $I_\mathrm{b}$ is the moment of inertia of the QCD vacuum energy, which is dominated by the gluon condensate~\cite{Shifman:1978bx,colangelo2001qcd} through the trace anomaly.
Therefore, for the rotational enhancement of $T_\mathrm{c}$, it is essential that the gluon condensate bears a large positive moment of inertia
(see Ref.~\cite{Braguta:2023yjn,Braguta:2023kwl} for a similar but different argument).
Such a situation would also be realized in lattice QCD simulations under rotation.

\section{Summary}
We analyzed the deconfinement transition of the bag model under rotation.
The finding is that the critical temperature is enhanced with angular velocity increased, as a result of the revolution of the confining potential, or more fundamentally, the moment of inertia of the gluon condensate.
This plausibly expounds the lattice QCD simulations indicating that rotation favors confinement.
In our model, we adopt a homogeneous bag distribution as the simplest case, which involves the minimal revolution effect. 
In realistic cases with an inhomogeneous bag distribution, hence, the confined phase should be more favored by rotation.
Our result also exhibits the importance of more rudimentary studies, such as the field-theoretical perturbation theory under rotation~\cite{Mameda:pre}.
Furthermore, nonperturbative analyses are ineludible in order to rigorously uncover the QCD vacuum property under rotation.

\section*{Acknowledgements}
The authors thank Yuki~Fujimoto, Xu-Guang~Huang, Takeshi~Morita, Artem~Roenko and Katsuhiko Suzuki for valuable comments.

\bibliography{rotation}

\begin{thebibliography}{10}
\expandafter\ifx\csname url\endcsname\relax
  \def\url#1{\texttt{#1}}\fi
\expandafter\ifx\csname urlprefix\endcsname\relax\def\urlprefix{URL }\fi
\expandafter\ifx\csname href\endcsname\relax
  \def\href#1#2{#2} \def\path#1{#1}\fi

\bibitem{STAR:2017ckg}
L.~Adamczyk, et~al., {Global $\Lambda$ hyperon polarization in nuclear
  collisions: evidence for the most vortical fluid}, Nature 548 (2017) 62--65.
\newblock \href {http://arxiv.org/abs/1701.06657} {\path{arXiv:1701.06657}},
  \href {http://dx.doi.org/10.1038/nature23004}
  {\path{doi:10.1038/nature23004}}.

\bibitem{Chen:2015hfc}
H.-L. Chen, K.~Fukushima, X.-G. Huang, K.~Mameda, {Analogy between rotation and
  density for Dirac fermions in a magnetic field}, Phys. Rev. D93~(10) (2016)
  104052.
\newblock \href {http://arxiv.org/abs/1512.08974} {\path{arXiv:1512.08974}},
  \href {http://dx.doi.org/10.1103/PhysRevD.93.104052}
  {\path{doi:10.1103/PhysRevD.93.104052}}.

\bibitem{Jiang:2016wvv}
Y.~Jiang, J.~Liao, {Pairing Phase Transitions of Matter under Rotation}, Phys.
  Rev. Lett. 117~(19) (2016) 192302.
\newblock \href {http://arxiv.org/abs/1606.03808} {\path{arXiv:1606.03808}},
  \href {http://dx.doi.org/10.1103/PhysRevLett.117.192302}
  {\path{doi:10.1103/PhysRevLett.117.192302}}.

\bibitem{Chernodub:2016kxh}
M.~N. Chernodub, S.~Gongyo, {Interacting fermions in rotation: chiral symmetry
  restoration, moment of inertia and thermodynamics}, JHEP 01 (2017) 136.
\newblock \href {http://arxiv.org/abs/1611.02598} {\path{arXiv:1611.02598}},
  \href {http://dx.doi.org/10.1007/JHEP01(2017)136}
  {\path{doi:10.1007/JHEP01(2017)136}}.

\bibitem{Chen:2022mhf}
Y.~Chen, D.~Li, M.~Huang, {Inhomogeneous chiral condensation under rotation in
  the holographic QCD}, Phys. Rev. D 106~(10) (2022) 106002.
\newblock \href {http://arxiv.org/abs/2208.05668} {\path{arXiv:2208.05668}},
  \href {http://dx.doi.org/10.1103/PhysRevD.106.106002}
  {\path{doi:10.1103/PhysRevD.106.106002}}.

\bibitem{Chen:2023cjt}
H.-L. Chen, Z.-B. Zhu, X.-G. Huang, {Quark-meson model under rotation: A
  functional renormalization group study}\href
  {http://arxiv.org/abs/2306.08362} {\path{arXiv:2306.08362}}.

\bibitem{Fujimoto:2021xix}
Y.~Fujimoto, K.~Fukushima, Y.~Hidaka, {Deconfining Phase Boundary of Rapidly
  Rotating Hot and Dense Matter and Analysis of Moment of Inertia}, Phys. Lett.
  B 816 (2021) 136184.
\newblock \href {http://arxiv.org/abs/2101.09173} {\path{arXiv:2101.09173}},
  \href {http://dx.doi.org/10.1016/j.physletb.2021.136184}
  {\path{doi:10.1016/j.physletb.2021.136184}}.

\bibitem{Chernodub:2020qah}
M.~N. Chernodub, {Inhomogeneous confining-deconfining phases in rotating
  plasmas}, Phys. Rev. D 103~(5) (2021) 054027.
\newblock \href {http://arxiv.org/abs/2012.04924} {\path{arXiv:2012.04924}},
  \href {http://dx.doi.org/10.1103/PhysRevD.103.054027}
  {\path{doi:10.1103/PhysRevD.103.054027}}.

\bibitem{Braga:2022yfe}
N.~R.~F. Braga, L.~F. Faulhaber, O.~C. Junqueira, {Confinement-deconfinement
  temperature for a rotating quark-gluon plasma}, Phys. Rev. D 105~(10) (2022)
  106003.
\newblock \href {http://arxiv.org/abs/2201.05581} {\path{arXiv:2201.05581}},
  \href {http://dx.doi.org/10.1103/PhysRevD.105.106003}
  {\path{doi:10.1103/PhysRevD.105.106003}}.

\bibitem{Chen:2020ath}
X.~Chen, L.~Zhang, D.~Li, D.~Hou, M.~Huang, {Gluodynamics and deconfinement
  phase transition under rotation from holography}, JHEP 07 (2021) 132.
\newblock \href {http://arxiv.org/abs/2010.14478} {\path{arXiv:2010.14478}},
  \href {http://dx.doi.org/10.1007/JHEP07(2021)132}
  {\path{doi:10.1007/JHEP07(2021)132}}.

\bibitem{Chen:2022smf}
S.~Chen, K.~Fukushima, Y.~Shimada, {Perturbative Confinement in Thermal
  Yang-Mills Theories Induced by Imaginary Angular Velocity}, Phys. Rev. Lett.
  129~(24) (2022) 242002.
\newblock \href {http://arxiv.org/abs/2207.12665} {\path{arXiv:2207.12665}},
  \href {http://dx.doi.org/10.1103/PhysRevLett.129.242002}
  {\path{doi:10.1103/PhysRevLett.129.242002}}.

\bibitem{Azuma:2023pzr}
T.~Azuma, T.~Morita, H.~Yoshida, {Complex Langevin Method on Rotating Matrix
  Quantum Mechanics at Thermal Equilibrium}\href
  {http://arxiv.org/abs/2302.14259} {\path{arXiv:2302.14259}}, \href
  {http://dx.doi.org/10.1093/ptep/ptad093} {\path{doi:10.1093/ptep/ptad093}}.

\bibitem{Yamamoto:2013zwa}
A.~Yamamoto, Y.~Hirono, {Lattice QCD in rotating frames}, Phys. Rev. Lett. 111
  (2013) 081601.
\newblock \href {http://arxiv.org/abs/1303.6292} {\path{arXiv:1303.6292}},
  \href {http://dx.doi.org/10.1103/PhysRevLett.111.081601}
  {\path{doi:10.1103/PhysRevLett.111.081601}}.

\bibitem{Braguta:2020biu}
V.~V. Braguta, A.~Y. Kotov, D.~D. Kuznedelev, A.~A. Roenko, {Study of the
  Confinement/Deconfinement Phase Transition in Rotating Lattice SU(3)
  Gluodynamics}, Pisma Zh. Eksp. Teor. Fiz. 112~(1) (2020) 9--16.
\newblock \href {http://dx.doi.org/10.31857/S1234567820130029}
  {\path{doi:10.31857/S1234567820130029}}.

\bibitem{Braguta:2021jgn}
V.~V. Braguta, A.~Y. Kotov, D.~D. Kuznedelev, A.~A. Roenko, {Influence of
  relativistic rotation on the confinement-deconfinement transition in
  gluodynamics}, Phys. Rev. D 103~(9) (2021) 094515.
\newblock \href {http://arxiv.org/abs/2102.05084} {\path{arXiv:2102.05084}},
  \href {http://dx.doi.org/10.1103/PhysRevD.103.094515}
  {\path{doi:10.1103/PhysRevD.103.094515}}.

\bibitem{Braguta:2021ucr}
V.~V. Braguta, A.~Y. Kotov, D.~D. Kuznedelev, A.~A. Roenko, {Lattice study of
  the confinement/deconfinement transition in rotating gluodynamics}, PoS
  LATTICE2021 (2022) 125.
\newblock \href {http://arxiv.org/abs/2110.12302} {\path{arXiv:2110.12302}},
  \href {http://dx.doi.org/10.22323/1.396.0125}
  {\path{doi:10.22323/1.396.0125}}.

\bibitem{Braguta:2022str}
V.~V. Braguta, A.~Kotov, A.~Roenko, D.~Sychev, {Thermal phase transitions in
  rotating QCD with dynamical quarks}, PoS LATTICE2022 (2023) 190.
\newblock \href {http://arxiv.org/abs/2212.03224} {\path{arXiv:2212.03224}},
  \href {http://dx.doi.org/10.22323/1.430.0190}
  {\path{doi:10.22323/1.430.0190}}.

\bibitem{Yang:2023vsw}
J.-C. Yang, X.-G. Huang, {QCD on Rotating Lattice with Staggered Fermions}\href
  {http://arxiv.org/abs/2307.05755} {\path{arXiv:2307.05755}}.

\bibitem{Vilenkin:1979ui}
A.~Vilenkin, Macroscopic parity violating effects: Neutrino fluxes from
  rotating black holes and in rotating thermal radiation, Phys. Rev. D20 (1979)
  1807--1812.
\newblock \href {http://dx.doi.org/10.1103/PhysRevD.20.1807}
  {\path{doi:10.1103/PhysRevD.20.1807}}.

\bibitem{Bali:2011qj}
G.~S. Bali, F.~Bruckmann, G.~Endrodi, Z.~Fodor, S.~D. Katz, S.~Krieg,
  A.~Schafer, K.~K. Szabo, {The QCD phase diagram for external magnetic
  fields}, JHEP 02 (2012) 044.
\newblock \href {http://arxiv.org/abs/1111.4956} {\path{arXiv:1111.4956}},
  \href {http://dx.doi.org/10.1007/JHEP02(2012)044}
  {\path{doi:10.1007/JHEP02(2012)044}}.

\bibitem{Bali:2012zg}
G.~S. Bali, F.~Bruckmann, G.~Endrodi, Z.~Fodor, S.~D. Katz, A.~Schafer, {QCD
  quark condensate in external magnetic fields}, Phys. Rev. D 86 (2012) 071502.
\newblock \href {http://arxiv.org/abs/1206.4205} {\path{arXiv:1206.4205}},
  \href {http://dx.doi.org/10.1103/PhysRevD.86.071502}
  {\path{doi:10.1103/PhysRevD.86.071502}}.

\bibitem{Fukushima:2012kc}
K.~Fukushima, Y.~Hidaka, {Magnetic Catalysis Versus Magnetic Inhibition}, Phys.
  Rev. Lett. 110~(3) (2013) 031601.
\newblock \href {http://arxiv.org/abs/1209.1319} {\path{arXiv:1209.1319}},
  \href {http://dx.doi.org/10.1103/PhysRevLett.110.031601}
  {\path{doi:10.1103/PhysRevLett.110.031601}}.

\bibitem{Fraga:2012fs}
E.~S. Fraga, L.~F. Palhares, {Deconfinement in the presence of a strong
  magnetic background: an exercise within the MIT bag model}, Phys. Rev. D 86
  (2012) 016008.
\newblock \href {http://arxiv.org/abs/1201.5881} {\path{arXiv:1201.5881}},
  \href {http://dx.doi.org/10.1103/PhysRevD.86.016008}
  {\path{doi:10.1103/PhysRevD.86.016008}}.

\bibitem{Kojo:2012js}
T.~Kojo, N.~Su, {The quark mass gap in a magnetic field}, Phys. Lett. B 720
  (2013) 192--197.
\newblock \href {http://arxiv.org/abs/1211.7318} {\path{arXiv:1211.7318}},
  \href {http://dx.doi.org/10.1016/j.physletb.2013.02.024}
  {\path{doi:10.1016/j.physletb.2013.02.024}}.

\bibitem{Kojo:2013uua}
T.~Kojo, N.~Su, {A renormalization group approach for QCD in a strong magnetic
  field}, Phys. Lett. B 726 (2013) 839--845.
\newblock \href {http://arxiv.org/abs/1305.4510} {\path{arXiv:1305.4510}},
  \href {http://dx.doi.org/10.1016/j.physletb.2013.09.023}
  {\path{doi:10.1016/j.physletb.2013.09.023}}.

\bibitem{Kugo:1977zq}
T.~Kugo, I.~Ojima, {Manifestly Covariant Canonical Formulation of Yang-Mills
  Field Theories: Physical State Subsidiary Conditions and Physical S Matrix
  Unitarity}, Phys. Lett. B 73 (1978) 459--462.
\newblock \href {http://dx.doi.org/10.1016/0370-2693(78)90765-7}
  {\path{doi:10.1016/0370-2693(78)90765-7}}.

\bibitem{Kugo:1979gm}
T.~Kugo, I.~Ojima, {Local Covariant Operator Formalism of Nonabelian Gauge
  Theories and Quark Confinement Problem}, Prog. Theor. Phys. Suppl. 66 (1979)
  1--130.
\newblock \href {http://dx.doi.org/10.1143/PTPS.66.1}
  {\path{doi:10.1143/PTPS.66.1}}.

\bibitem{Gribov:1977wm}
V.~N. Gribov, {Quantization of Nonabelian Gauge Theories}, Nucl. Phys. B 139
  (1978) 1.
\newblock \href {http://dx.doi.org/10.1016/0550-3213(78)90175-X}
  {\path{doi:10.1016/0550-3213(78)90175-X}}.

\bibitem{Zwanziger:1993dh}
D.~Zwanziger, {Fundamental modular region, Boltzmann factor and area law in
  lattice gauge theory}, Nucl. Phys. B 412 (1994) 657--730.
\newblock \href {http://dx.doi.org/10.1016/0550-3213(94)90396-4}
  {\path{doi:10.1016/0550-3213(94)90396-4}}.

\bibitem{Chodos:1974je}
A.~Chodos, R.~L. Jaffe, K.~Johnson, C.~B. Thorn, V.~F. Weisskopf, {A New
  Extended Model of Hadrons}, Phys. Rev. D 9 (1974) 3471--3495.
\newblock \href {http://dx.doi.org/10.1103/PhysRevD.9.3471}
  {\path{doi:10.1103/PhysRevD.9.3471}}.

\bibitem{Chodos:1974pn}
A.~Chodos, R.~L. Jaffe, K.~Johnson, C.~B. Thorn, {Baryon Structure in the Bag
  Theory}, Phys. Rev. D 10 (1974) 2599.
\newblock \href {http://dx.doi.org/10.1103/PhysRevD.10.2599}
  {\path{doi:10.1103/PhysRevD.10.2599}}.

\bibitem{Bhaduri:1988gc}
R.~K. Bhaduri, Models of the Nucleon: From Quark to Soliton, 1988.

\bibitem{greiner2007quantum}
W.~Greiner, S.~Schramm, E.~Stein, Quantum chromodynamics, Springer Science \&
  Business Media, 2007.

\bibitem{Jiang:2021izj}
Y.~Jiang, {Chiral vortical catalysis}, Eur. Phys. J. C 82~(10) (2022) 949.
\newblock \href {http://arxiv.org/abs/2108.09622} {\path{arXiv:2108.09622}},
  \href {http://dx.doi.org/10.1140/epjc/s10052-022-10915-8}
  {\path{doi:10.1140/epjc/s10052-022-10915-8}}.

\bibitem{Davies:1996ks}
P.~C.~W. Davies, T.~Dray, C.~A. Manogue, {The Rotating quantum vacuum}, Phys.
  Rev. D53 (1996) 4382--4387.
\newblock \href {http://arxiv.org/abs/gr-qc/9601034}
  {\path{arXiv:gr-qc/9601034}}, \href
  {http://dx.doi.org/10.1103/PhysRevD.53.4382}
  {\path{doi:10.1103/PhysRevD.53.4382}}.

\bibitem{Ambrus:2015lfr}
V.~E. Ambrus, E.~Winstanley, {Rotating fermions inside a cylindrical boundary},
  Phys. Rev. D93~(10) (2016) 104014.
\newblock \href {http://arxiv.org/abs/1512.05239} {\path{arXiv:1512.05239}},
  \href {http://dx.doi.org/10.1103/PhysRevD.93.104014}
  {\path{doi:10.1103/PhysRevD.93.104014}}.

\bibitem{Ebihara:2016fwa}
S.~Ebihara, K.~Fukushima, K.~Mameda, {Boundary effects and gapped dispersion in
  rotating fermionic matter}, Phys. Lett. B 764 (2017) 94--99.
\newblock \href {http://arxiv.org/abs/1608.00336} {\path{arXiv:1608.00336}},
  \href {http://dx.doi.org/10.1016/j.physletb.2016.11.010}
  {\path{doi:10.1016/j.physletb.2016.11.010}}.

\bibitem{rose1995elementary}
M.~E. Rose, Elementary theory of angular momentum, John Wiley \& Sons, New
  York, 1995.

\bibitem{Zhang:2020hct}
Z.~Zhang, C.~Shi, X.~Luo, H.-S. Zong, {Rotating fermions inside a spherical
  boundary}, Phys. Rev. D 102~(6) (2020) 065002.
\newblock \href {http://arxiv.org/abs/2006.00677} {\path{arXiv:2006.00677}},
  \href {http://dx.doi.org/10.1103/PhysRevD.102.065002}
  {\path{doi:10.1103/PhysRevD.102.065002}}.

\bibitem{Aidala:2012mv}
C.~A. Aidala, S.~D. Bass, D.~Hasch, G.~K. Mallot, {The Spin Structure of the
  Nucleon}, Rev. Mod. Phys. 85 (2013) 655--691.
\newblock \href {http://arxiv.org/abs/1209.2803} {\path{arXiv:1209.2803}},
  \href {http://dx.doi.org/10.1103/RevModPhys.85.655}
  {\path{doi:10.1103/RevModPhys.85.655}}.

\bibitem{Hattori:2016njk}
K.~Hattori, Y.~Yin, {Charge redistribution from anomalous magnetovorticity
  coupling}, Phys. Rev. Lett. 117~(15) (2016) 152002.
\newblock \href {http://arxiv.org/abs/1607.01513} {\path{arXiv:1607.01513}},
  \href {http://dx.doi.org/10.1103/PhysRevLett.117.152002}
  {\path{doi:10.1103/PhysRevLett.117.152002}}.

\bibitem{Chen:2017xrj}
H.-L. Chen, K.~Fukushima, X.-G. Huang, K.~Mameda, {Surface Magnetic Catalysis},
  Phys. Rev. D 96~(5) (2017) 054032.
\newblock \href {http://arxiv.org/abs/1707.09130} {\path{arXiv:1707.09130}},
  \href {http://dx.doi.org/10.1103/PhysRevD.96.054032}
  {\path{doi:10.1103/PhysRevD.96.054032}}.

\bibitem{Savvidy:1977as}
G.~K. Savvidy, {Infrared Instability of the Vacuum State of Gauge Theories and
  Asymptotic Freedom}, Phys. Lett. B 71 (1977) 133--134.
\newblock \href {http://dx.doi.org/10.1016/0370-2693(77)90759-6}
  {\path{doi:10.1016/0370-2693(77)90759-6}}.

\bibitem{Nielsen:1978rm}
N.~K. Nielsen, P.~Olesen, {An Unstable Yang-Mills Field Mode}, Nucl. Phys. B
  144 (1978) 376--396.
\newblock \href {http://dx.doi.org/10.1016/0550-3213(78)90377-2}
  {\path{doi:10.1016/0550-3213(78)90377-2}}.

\bibitem{Nielsen:1978tr}
H.~B. Nielsen, M.~Ninomiya, {A Bound on Bag Constant and Nielsen-Olesen
  Unstable Mode in QCD}, Nucl. Phys. B 156 (1979) 1--28.
\newblock \href {http://dx.doi.org/10.1016/0550-3213(79)90490-5}
  {\path{doi:10.1016/0550-3213(79)90490-5}}.

\bibitem{huang1992quarks}
K.~Huang, Quarks, leptons \& gauge fields, World Scientific, Singapore, 1992.

\bibitem{Shifman:1978bx}
M.~A. Shifman, A.~I. Vainshtein, V.~I. Zakharov, {QCD and Resonance Physics.
  Theoretical Foundations}, Nucl. Phys. B 147 (1979) 385--447.
\newblock \href {http://dx.doi.org/10.1016/0550-3213(79)90022-1}
  {\path{doi:10.1016/0550-3213(79)90022-1}}.

\bibitem{colangelo2001qcd}
P.~Colangelo, A.~Khodjamirian, {QCD sum rules, a modern perspective}, in:
  M.~Shifman, B.~Ioffe (Eds.), At The Frontier of Particle Physics: Handbook of
  QCD (in 3 Volumes), World Scientific, Singapore, 2001, pp. 1495--1576.
\newblock \href {http://arxiv.org/abs/hep-ph/0010175}
  {\path{arXiv:hep-ph/0010175}}, \href
  {http://dx.doi.org/10.1142/9789812810458_0033}
  {\path{doi:10.1142/9789812810458_0033}}.

\bibitem{Braguta:2023yjn}
V.~V. Braguta, M.~N. Chernodub, A.~A. Roenko, D.~A. Sychev, {Negative moment of
  inertia and rotational instability of gluon plasma}\href
  {http://arxiv.org/abs/2303.03147} {\path{arXiv:2303.03147}}.

\bibitem{Braguta:2023kwl}
V.~V. Braguta, I.~E. Kudrov, A.~A. Roenko, D.~A. Sychev, M.~N. Chernodub,
  {Lattice Study of the Equation of State of a Rotating Gluon Plasma}, JETP
  Lett. 117~(9) (2023) 639--644.
\newblock \href {http://dx.doi.org/10.1134/S0021364023600830}
  {\path{doi:10.1134/S0021364023600830}}.

\bibitem{Mameda:pre}
R.~Kuboniwa, K.~Mameda, {in preparation}.

\end{thebibliography}
\bibliographystyle{elsarticle-num}

\end{document}